\documentclass[
twocolumn,preprintnumbers,
amsmath, 
amssymb, 
aps, 
nofootinbib,
raggedbottom,
superscriptaddress]{revtex4-1}
\pdfoutput=1

\usepackage{graphicx}
\usepackage{enumitem}

\usepackage{hyperref}

\begin{document}

\preprint{LAPTH-005/24}

\title{The Regge bootstrap, from linear to non-linear trajectories}

\author{Christopher Eckner}
\email{eckner@lapth.cnrs.fr}
\affiliation{LAPTh, CNRS et Universit\'e Savoie Mont-Blanc, 9 Chemin de Bellevue, F-74941 Annecy, France}
\affiliation{LAPP, CNRS et Universit\'e Savoie Mont-Blanc, 9 Chemin de Bellevue, F-74941 Annecy, France}
\affiliation{Center for Astrophysics and Cosmology, University of Nova Gorica, Vipavska 11c, 5270 Ajdov\v{s}\v{c}ina,
Slovenia}

\author{Felipe Figueroa}
\email{figueroa@lapth.cnrs.fr}
\affiliation{LAPTh, CNRS et Universit\'e Savoie Mont-Blanc, 9 Chemin de Bellevue, F-74941 Annecy, France}

\author{Piotr Tourkine}
\email{tourkine@lapth.cnrs.fr}
\affiliation{LAPTh, CNRS et Universit\'e Savoie Mont-Blanc, 9 Chemin de Bellevue, F-74941 Annecy, France}

\begin{abstract}

We present a numerical linear programming bootstrap to construct dual model scattering amplitudes. Dual models describe tree-level exchanges of higher spin resonances in theories like string theory and large $N$ gauge theories. Despite being very simple objects, their numerical bootstrap has proven challenging due to slow convergence of the infinite sums over resonances.
Our bootstrap succeeds thanks to an efficient parametrization of the amplitude in terms of Mandelstam-Regge poles and the use of combined regions that make crossing symmetry constraining. Along the way, we discover and conjecture a property of ``super-unitarity'' of the Veneziano amplitude, which we use to keep a linear problem.

As results, we present first the study of a class of string-like amplitudes with linear trajectories, for which we observe that the Veneziano amplitude lies at a preferred location, at the bottom of a pit, which minimizes crossing.
Then, we introduce a toy-model deformation to non-linear trajectories, mimicking some features of QCD, for which our algorithm also detects a clear pit.
This gives compelling evidence that our bootstrap is able to produce amplitudes that can exhibit non-trivial phenomenological features.

\end{abstract}

\maketitle

\section{Introduction}

Dual models have a long history. Originally designed to understand the strong force, they gave rise to the celebrated Veneziano amplitude~\cite{Veneziano:1968yb} and ultimately string theory. 
They describe the exchange of infinitely many higher spin massive resonances at tree-level, having only single poles as singularities. Therefore they describe weakly coupled phenomena, such as tree-level string theory and large-$N$ gauge theories, yet they can probe non-perturbative effects, such as confinement in gauge theories. 
Being meromorphic, they are made of single poles and their residues. They are very simple-looking objects and the simplest non-trivial scattering amplitudes that one can hope to build explicitly. 

Starting with~\cite{Caron-Huot:2016icg}, recent years have seen remarkable progress happen on the analytical side (see discussion section and in particular the constructions of~\cite{Cheung:2023uwn} and~\cite{Haring:2023zwu}). In~\cite{Haring:2023zwu}, the authors used  powerful primal and dual bootstrap tools for amplitudes with linear, integer-spaced spectra. However, and despite the existence of many tools for the full non-perturbative $S$-matrix~\cite{Kruczenski:2022lot}, at the present time, a scheme to construct generic dual model amplitudes with an arbitrary spectrum does not exist.
The main reason why the problem is difficult is because the infinite sum over the resonances converges very slowly, thus a truncated amplitude does not approximate the amplitude well, therefore crossing cannot be constraining.

\begin{figure}[h]
    \centering
    \includegraphics[width=0.95\columnwidth]{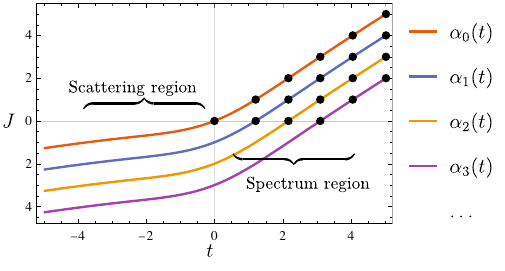} 
    \caption{Example of bent Regge trajectories for which we can construct amplitudes in dimensionless units $\alpha'=1$.}
    \label{fig:bent-trajectories}
\end{figure}

Because of this difficult numerical challenge, several open questions remain unknown about these meromorphic amplitudes: What is the space of dual models? Can we construct amplitudes of phenomenological interest, e.g., for strong interactions or holography? In this context, it is particularly important to build amplitudes with non-linear Regge trajectories $\alpha(t)$ that depart from the standard linear behavior of the Veneziano amplitude at negative $t$, as in fig.~(\ref{fig:bent-trajectories}), as this is the expected behavior to reproduce the high energy scattering of hadrons in QCD.

In this paper, we construct a numerical bootstrap scheme for dual models with linear or non-linear trajectories.
We combine the standard assumptions of analyticity, crossing and unitarity (ACU) within a linear programming paradigm granting us high computational speed.

Our bootstrap takes as an \textit{input} the \textit{set} of Regge trajectories $\alpha_i(t)$, from the leading one $i=0$ to a maximum cut-off  $i=N_{\text{traj}}$, and a tolerance parameter which indicates by how much we demand crossing to be satisfied. It produces as an \textit{output} the Regge residues $\beta_i(t)$ (introduced below). Combined with the Regge trajectories, these data allow us to fully reconstruct the amplitude.

The three main novelties that allow our bootstrap to work are: (i) the use of the Mandelstam-Regge (MR) pole expansion, which allows for repackaging the infinity of couplings on a given trajectory into a function with a finite amount of parameters;
(ii)  the determination of regions where crossing symmetry can be applied which are sensitive to both the spectrum and scattering part of the amplitude ($t>0$ and $t<0$, respectively, see fig.~\ref{fig:bent-trajectories}), and (iii) the use of a "super-unitarity" constraint, inspired from string theory, which ensures the linearity of the problem.

Our bootstrap is constructive, and \textit{a priori} we cannot say if an output amplitude, for a given tolerance parameter, will truly converge or not.
Therefore, in order to be able to interpret the output of the bootstrap, we calibrate it empirically on a restricted, but non-trivial polynomial ansatz, inspired by the Veneziano amplitude, in sec.~\ref{sec:results-vene}. This teaches us that Veneziano lies at a preferred location, easily identifiable by a large drop in our tolerance parameter. This test is a non-trivial confirmation of our numerical strategy, because any naive approach simply does not produce an amplitude at all.

Armed with the knowledge that our algorithm detects amplitudes via a dip, we proceed to deform the trajectories to phenomenological features (see fig.~\ref{fig:bent-trajectories}) defined by two different slopes in the spectrum and scattering regions, motivated by the fact that in Quantum Field Theory we expect the slope on the scattering region to go to zero, up to logarithmic corrections.

By tuning the slope, we generate new amplitudes -- deformations of the Veneziano amplitude -- which we identify through the large drop in the tolerance parameter, strongly hinting at the existence of such objects within the paradigm of this bootstrap. We can track and display various observables across the deformation.

\section{Setup}

In this paper, we study the $2\to2$ scattering of identical, color-ordered scalars.
The prototypical example would be a single Veneziano block\footnote{In contrast, a non-color ordered scalar amplitude would be given by $V(s,t)+V(s,u)+V(t,u)$.}
\begin{equation}
    V(s,t)=-\frac{\Gamma(m^2-s)\Gamma(m^2-t)}{\Gamma(2m^2-s-t)}\rm{.}
\end{equation}
For simplicity, we choose the scalars to be massless, $m^2=0$ and we choose to work in $d=4$, though both conditions can be relaxed. Our assumptions are as follows: 

\paragraph{Analyticity, locality, maximal spin.}
We work with meromorphic amplitudes, hence the only singularities are single poles, corresponding to the real masses of the stable resonances. 

Besides, we work with dual amplitudes, that have by definition the property that the sum of poles in one channel produces the singularities of the other channel.\footnote{See some recent developments in~\cite{Saha:2024qpt}.} 
A standard fixed-$t$ dispersion relation gives 
\begin{equation}
    \label{eq:DR}
    A(s,t) = \frac1{2i\pi}\oint \frac{A(s',t)}{s'-s}\mathrm{d}s^{\prime}=\sum_{n=0}^\infty \frac{{\rm Res}(A(s,t))|_{s=m_n^2}}{s-m_n^2}\rm{.}
\end{equation}
For $t<m_0^2$, the amplitudes decay at infinity so that the arc at infinity of the dispersion relation can be dropped.\footnote{The question of the asymptotic behavior of amplitudes has been investigated in~\cite{Caron-Huot:2016icg} and recently in~\cite{Haring:2023zwu} where Regge-sum-rules were used to constrain amplitudes with linear spectra in particular.}
The series diverges as soon as $t$ hits the first singularity, $t=m_0^2$.

\textit{Locality} here requires to have polynomial residues, that can be expanded into the standard Gegenbauer polynomial basis, 
so that the form of the amplitude we work with is
\begin{equation}
    \label{eq:duality}
    A(s,t) = \sum_{n=0}^\infty \sum_{J=0}^n \frac{c_{n,J}P_J(1+\frac{2t}{m_n^2-4m_0^2})}{s-m_n^2}\rm{.}
\end{equation}
We also use an assumption of "Maximal spin", as in \cite{Caron-Huot:2016icg,Haring:2023zwu}, which for us is that the degree of ${\rm Res}(A(s,t))|_{s=m_n^2}$ is at most the level $n$.

\paragraph{Unitarity.}
Meromorphic amplitudes satisfy tree-level unitarity, which boils down to a positivity condition
\begin{equation}
    c_{n,J}\geq0
\end{equation}
that ensures that no states with negative norm are being exchanged.

\paragraph{Crossing,}
\begin{equation}
    A(s,t) = A(t,s)\,,
\end{equation}
is the most delicate constraint to enforce in this construction. With a finite, truncated ansatz, crossing can never be exact. Therefore, we apply it on a grid of points with a tolerance factor $\epsilon$, and we ask that for an amplitude to be possibly physical, as we increase the number of trajectories, this tolerance decreases.

\paragraph{Reggeization.}
For our bootstrap to work, we assume that the amplitude reggeizes, and we explicitly parametrize the amplitude as a sum of MR poles.

\section{The Mandelstam-Regge Bootstrap}

We can now present our bootstrap, firstly describing how we parametrize our ansatz based on the MR expansion and then explaining how we impose unitarity and crossing symmetry.

\subsection{Mandelstam-Regge pole expansion (MRPE)}
\label{sec:MRPE}

As an ansatz for the scattering amplitude, we use the expansion over Mandelstam-Regge poles~\cite{MANDELSTAM1962254}, truncated at some finite number of trajectories $N_{\text{traj}}$. 
In the $s$-channel, in $d=4$, this expansion reads
\begin{equation}
\label{eq:truncatedMRexp}
 A^{N_{\text{traj}}}(s,t)=  16 \pi^2 \sum_{i=0}^{N_\text{traj}}(2\alpha_i(s)+1)\beta_i (s)\frac{Q_{-\alpha_i (s)-1}(-z)}{\cos \pi \alpha_i (s)},
\end{equation}
with $z=1+\frac{2t}{s}$ and the Legendre $Q$ function defined in appendix~\ref{app:definitions}

This sum is expected to be asymptotic, and in the language of Regge theory, retaining only the poles amounts to neglecting the contribution of the background integral in the Sommerfeld-Watson transform. 
Mandelstam's refinement above was introduced to improve the standard Regge pole expansion and make the background terms small in a large region of the complex plane. 
Consequently, a remarkable feature of this ansatz is that -- despite being an asymptotic expansion -- including only a few terms in the sum allows for approximating the full amplitude to very high accuracy. We refer to appendix~\ref{app:MRP} for more details.%

Once the $\alpha_i(s)$ are fixed, all terms except the $\beta_i(s)$'s in eq.~(\ref{eq:truncatedMRexp}) are given by kinematics. Hence, the $\beta_i$'s completely define the amplitude. This can also be seen by relating the MR poles to the dispersion relation~\eqref{eq:DR}. Indeed, the spectrum and the couplings are completely determined by the Regge trajectories and Regge residues via
\begin{equation}
   c_{n,n-i} = 16 \pi^2 (2n-2i+1)\beta_i(m_n^2) , \quad m_{n,n-i}^2=\alpha_i^{-1}(n),
    \label{eq:beta-cnJ}
\end{equation}
and thus the full Regge expansion carries the same information as eq.~(\ref{eq:duality}). We give a few details on these standard facts in appendix~\ref{app:regge-vs-DR}.

Another nice feature of this ansatz is that it guarantees \textit{by construction} the polynomiality of the residues.

Our exact ansatz is finally defined by selecting a parametrization for the functions $\beta_i$. For this, we use a polynomial parametrization crucially inspired by the Veneziano amplitude, for which we derived that residues take the following form~(some details are given in app.~\ref{app:vene-residues}): 
\begin{equation}
    \label{eq:beta_ansatz}
    \beta^{(V)}_j (s) =B(s) \frac{1}{s^{d_j}}p^{(V)}_j (s), \qquad p^{(V)}_j (s)=\sum_{j=0}^{q_j} a^{(V)}_{jk}s^k
\end{equation}
where 
\begin{equation}
    \label{eq:UniversalFunctionVeneziano}
    B(s)= \frac{\pi ^{\frac{1}{2}-\frac{d}{2}} s^s 2^{-2 d-2 s+3}}{\Gamma \left(\frac{d-1}{2}+s\right)}\,
\end{equation}
is a universal factor that does not depend on the trajectory, and we observed empirically by computing over 30 residues that $d_0=0$, $d_1=1$, $d_{j\geq2} = 2\lfloor \frac{j}2\rfloor$, and  $p^{(V)}_j$ is a polynomial of degree $q_j$ with $q_0=0$, $q_1=1$, $q_{j\geq2}=3\lfloor \frac{j}2\rfloor$. Note that, $j=0$ corresponds to the \textit{first} trajectory.

For our final ansatz, we keep the universal function, which we simply modify by replacing $s$ by $\alpha(s)$,
so that our ansatz is, explicitly:
\begin{equation}
\beta_j^{\rm(our~ansatz)} =  B(\alpha(s))\frac{p_j (s)}{s^{d_j}}, \qquad p_j (s)=\sum_{j=0}^{q_j} a_{jk}s^k
\label{eq:beta-ansatz}
\end{equation}
where the parameters of our ansatz are the constants $a_{ij}$ which are now free parameters. Through eq.~\eqref{eq:beta-cnJ}, the $a$'s are linearly related to the $c_{n,J}$, see a few examples in eq.~\eqref{eq:a-to-c}.

While the number of parameters grows like
\begin{equation}
    \# {\rm params}\sim \sum_{i=0}^{N_{\rm traj}} i \sim N_{\rm traj}^2\mathrm{.}
\end{equation}
the efficiency of our approach lies in the fact that each trajectory comprises the information of infinitely many coefficients $c_{n,J}$ and thus convergence can occur very fast.

One dangerous feature of the MRPE is that the $(\cos{\pi \alpha_i})^{-1}$ terms introduce an infinite number of spurious poles at half-integer values of the $\alpha_i(s)$'s.\footnote{The authors of \cite{Mandelstam:1968zza,Dietz:1968pz} attempted to engineer a bootstrap of the trajectory functions themselves, by requiring the residues to cancel these poles.} In the case of Veneziano, the poles at $s<-1/2$ are canceled by the zeros of $B(s)$, and so are they in our modified ansatz. The zero at $s=-1/2$ instead is removed by the polynomials $p_j$, which all of them have a zero at that value. Inspired by this remarkable feature, we impose this behavior at $s=-1/2$ on our ansatz, too. In contrast, the poles at positive semi-integers are not removed and thus spoil the convergence of the MR expansion on the positive real axis. As one departs a little from the real line the expansion quickly becomes very accurate again, as shown in fig.~\ref{fig:MR-vs-Vene}.

\subsection{Unitarity; super-unitarity}
\label{sec:unitarity-super}
From eq.~(\ref{eq:beta-cnJ}) we see that in terms of the MR expansion unitarity reduces to the statement that the residue functions are positive at $s=m_n^2$
\begin{equation}
    {\rm Unitarity\,:\quad}\forall n,i,\,\beta_i(m_n^2) \geq0\rm{.}
\end{equation}
Since $B(s)$ is positive for $s\geq0$, unitarity becomes the condition that the polynomials $p_i(s)$ in the ansatz are positive at $s=m_n^2$ for all $n$. 

However, we observed that the Veneziano amplitude satisfies an even stronger property that we refer to as {\it super-unitarity}: the shifted polynomials $p^{(V)}_j(x+j)$, which should, by what we just said, be positive on positive integers $x=n\geq0$, are actually polynomials \textit{with only positive coefficients}.
\footnote{This pattern is true for all the Regge residues we computed, up to the 37th trajectory in $d=4$ and in all dimensions $d\leq10$ up to the 15th trajectory. Proving it in general offers a possibly simpler new avenue for showing directly the positivity of the Veneziano amplitude from its explicit expression~\cite{Arkani-Hamed:2022gsa}.} It therefore seems that string theory could be \textit{more} unitary than it needs to be, and raises the question of whether there exists an even more unitary amplitude which saturates ACU bounds.

For our bootstrap, this super-unitarity allows us to keep a linear problem, so we use it and impose: 
\begin{equation}
\label{eq:positivityofcoefs}
    p_i(x+\alpha_i^{-1}(0)) = \sum_{j=0}^{q_i} b_{ij} x^j ,\quad b_{ij}\geq0.
\end{equation}
In this way, each $b_{ij}$ is a linear combination of the original $a_{ij}$ coefficients.

\subsection{Crossing}
The final and most delicate aspect of our bootstrap procedure is to impose crossing on our MR ansatz. It cannot be satisfied exactly with a truncated ansatz and hence we need to impose it {\it approximately}. Moreover, it is necessary to determine the regions in $s,t$ where the MR expansion in a given channel provides a good approximation for the full amplitude, as it is reasonable to impose crossing only on the intersection between these regions for both the $s$ and $t$-channels. 

We address the first issue by demanding that for a given $N_{\text{traj}}$ the ansatz satisfies crossing up to a \textit{uniform tolerance} $\epsilon$. To be precise, we impose
\begin{equation}
    \label{eq:crossingconstraints}
   \left|1-\frac{A^{N_\text{traj}}(t_j,s_j)}{A^{N_\text{traj}}(s_j,t_j)}\right|\leq \epsilon.
\end{equation}
on a set of points $\{(s_j,t_j)\}_{j=1}^N$ made of distinct regions detailed below.\footnote{We thank Denis Karateev for introducing us to this concept of tolerance.}
%
When $s_j,t_j\in \mathbb{R}$, assuming a sign for $A^{N_\text{traj}}(s_j,t_j)$ allows us to reformulate eq.~(\ref{eq:crossingconstraints}) as a linear inequality invoking the ansatz parameters, for instance, if $A^{N}(s_j,t_j)\geq0$,
\begin{equation}
    \label{eq:crossingconstraints-linear}
   -\epsilon{A^{N}(s_j,t_j)}\leq {A^{N}(s_j,t_j)}-{A^{N}(t_j,s_j)}\leq \epsilon{A^{N}(s_j,t_j)}.
\end{equation}
For the other sign, the inequality is reversed. 
Importantly, note that when $s,t$ are complex, we restrict ourselves to applying crossing on the \textit{real part} only.

Sampling over a suitable set of values for $s,t$ then generates the constraints that we use to impose crossing on the amplitude.

Understanding in which regions of $(s,t)\in\mathbb{C}^2$ we can impose crossing required a detailed study of the regions of convergence and the analytical structure of the MR expansion. While we relegate the details of this analysis to the appendix~\ref{app:crossing}, our main finding, which is instrumental in making this bootstrap work, is that crossing can be enforced on points lying on rays satisfying $\Im(s/t)=0$ (together with a particular $i \epsilon$ prescription), and far enough from the singularities of the $Q_J$ functions at $s=-t$, $s=0$ and $t=0$ as well as the positive real axis, where the expansion breaks down.

In practice we impose crossing on a grid of points in four different regions of $(s,t)\in\mathbb{C}^2$ shown in fig.~(\ref{fig:schematic-constraints}): R1 = Real, negative $s,t$; R2 = Set of rays $\Im s/t=0$ with $\theta \in (2.9,3.4)$; R3 = Point $(4 e^{i/4},5 e^{i/4})$; R4 = Point $(\frac{9}{2} e^{i6/5},2 e^{i6/5})$.

It is crucial to emphasize that, although imposing crossing solely in the region of $\Im(s,t)\leq0$ appeared to suffice for linear trajectories, albeit with a slower convergence rate, the inclusion of points in the positive imaginary part region is a pivotal factor for investigating amplitudes associated with bending trajectories, for otherwise the crossing equation is never evaluated in the region where $\alpha(t)$ starts to differ from its value at $t<0$. Empirically, we observe that introducing even a single point in this region is adequate to achieve converged bounds.

Finally, note that the question of the sign used to define~\eqref{eq:crossingconstraints-linear} is important: as we observed on the MRPE of the Veneziano amplitude, it is mostly determined by the sign of the leading trajectory, which depends only on the universal function as the polynomial is trivial. Hence, we could choose the sign beforehand to be that of the leading trajectory. In practice, we choose most of the points in the $s,t<0$, where the Veneziano amplitude and leading trajectories do not change sign. There we can impose consistently the same sign. 
In the other regions (such as $R_3$ in fig.\ref{fig:schematic-constraints}), the amplitude can change sign and different signs will lead to different amplitudes. In the paper we adopt the Veneziano sign for the few points which we take in the region where the sign is free, in appendix \ref{app:sign-flip} we display the results when we select a different sign and explore some properties of the discovered object (see fig.~\ref{fig:eft_ellipses}).

\subsection{Observables}
To describe the space of amplitudes in our class we must extremize some observables that depend on the ansatz's parameters. A standard choice are the Wilson coefficients determining the effective field theory expansion of the amplitude, but this is not easily accessible for our MR ansatz, because the $Q_J$ functions have logarithmic singularities at $s,t=0$, preventing a well-defined Taylor expansion around this point. Defining them via dispersion relations also fails, as the contours of integration pass through regions of $\mathbb{C}^2$ where the MR expansion does not match the amplitude. 

Hence, to parametrize our space of amplitudes, we use another class of observables: the value of the amplitude and its derivatives near a point where the MR expansion is well convergent. We choose an arbitrary, negative, rational point which is neither integer nor half-integer, $s,t=-5/3$ in dimensionless units, and define:
\begin{equation}
\label{eq:defgijs}
g_{ij}=\frac{\partial^{i+j}}{\partial s^i \partial t^j}A(s,t)\bigg|_{s=t=-5/3}.
\end{equation}
Since these observables\footnote{It would be interesting to derive the space of such couplings from standard null-constraint~\cite{Caron-Huot:2020cmc,Arkani-Hamed:2020blm,Bellazzini:2020cot,Tolley:2020gtv,Bern:2021ppb,deRham:2022hpx} analysis, to i) check if, and ii) see where, our amplitudes lie within this allowed space.} are linear functions of the ansatz parameters, imposing crossing and unitarity remains a linear optimization problem over the $\{a_{ij}\}$. Finally, since for us unitarity is only positivity, we also need to set a scale for the amplitude, and we impose $g_{00}=1$.

\section{Results}
\label{sec:results}

\begin{figure}
    \centering
    \includegraphics[width=\columnwidth]{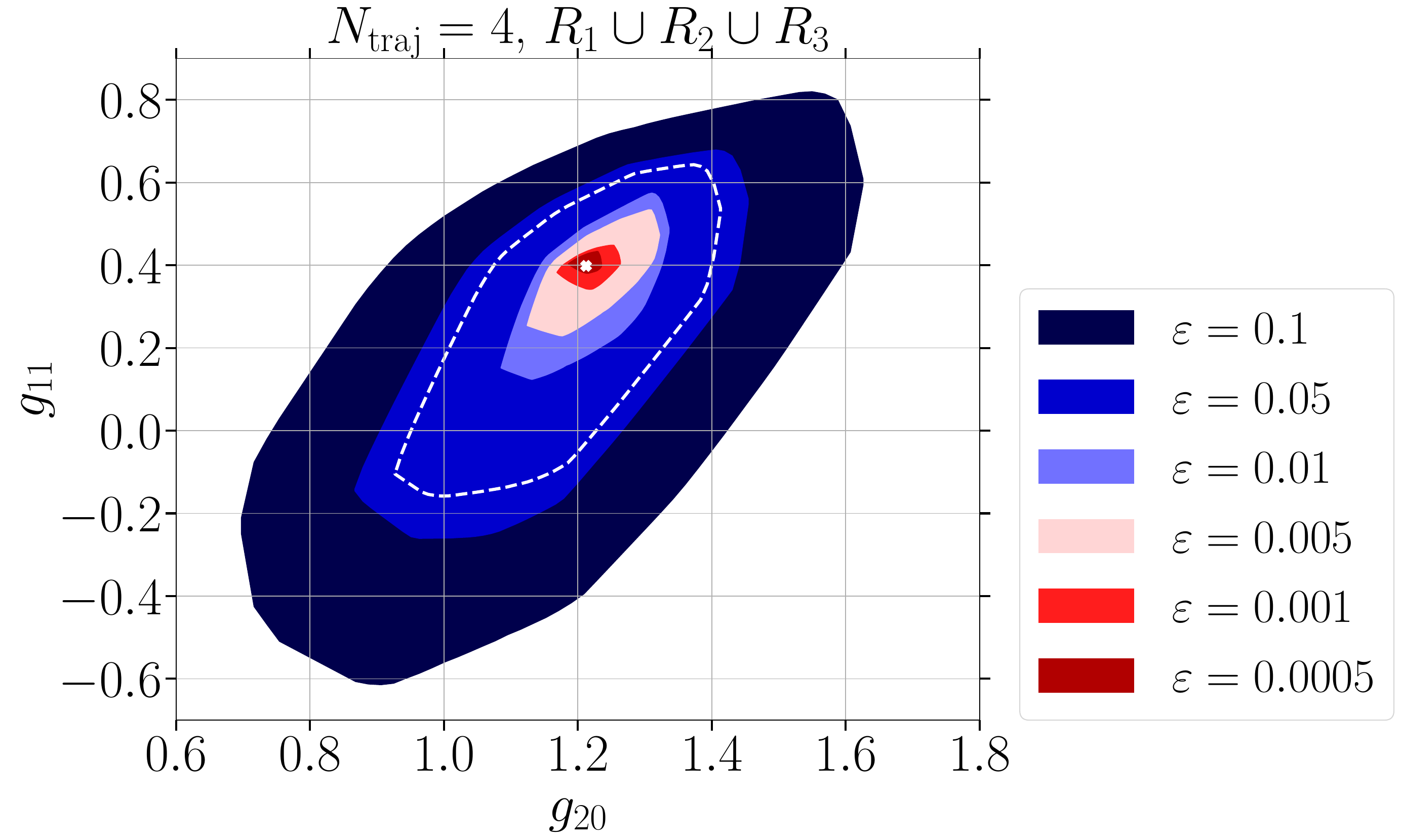}
    \caption{
     Linear trajectories ansatz. Allowed space in the $\left(g_{20}, g_{11}\right)$-plane evaluated at the point $(s, t) = (-5/3, -5/3)$ traced for decreasing values of the tolerance parameter $\epsilon$. We show the results for a Regge ansatz of $N_{\mathrm{traj}}=5$ and imposed crossing constraints from the set $R_1 \cup R_2 \cup R_3$. The white cross refers to the value of the Veneziano amplitude in this parameter space recaled to match our requirement of $g_{00} = 1$. The dashed, white line indicates the minimal tolerance $\epsilon = 3.65\%$ that the MRPE achieves when inserting the $\{a_{ij}\}$ of the Veneziano amplitude. Here, we find that the Veneziano amplitude is detected by our bootstrap approach.}
     \label{fig:shrinking-ellipses-vene}
\end{figure}

We now display numerical results from implementing this algorithm. We give details on the numerics in the discussion and the appendix, but let us just mention that the results presented here were obtained with \textsc{Mosek}~\cite{andersen2000mosek}, a very powerful linear programming solver.

In the first part of this section, we focus on the case with linear, integer-spaced trajectories. The main result is a result about the bootstrap itself: the tolerance parameter suddenly drops significantly at the known location of the Veneziano amplitude.
Then we move in to a class of models with bent trajectories, to demonstrate that it can continue to generate non-trivial amplitudes.

\subsection{Pure linear trajectories}
\label{sec:results-vene}
Here we apply the bootstrap explained above with $\alpha_i(s) = \alpha's-i$, and work with $\alpha'=1$ units. The corresponding polynomial ansatz is still non-trivial. 

\begin{figure}
    \centering
    \includegraphics[width=\columnwidth]{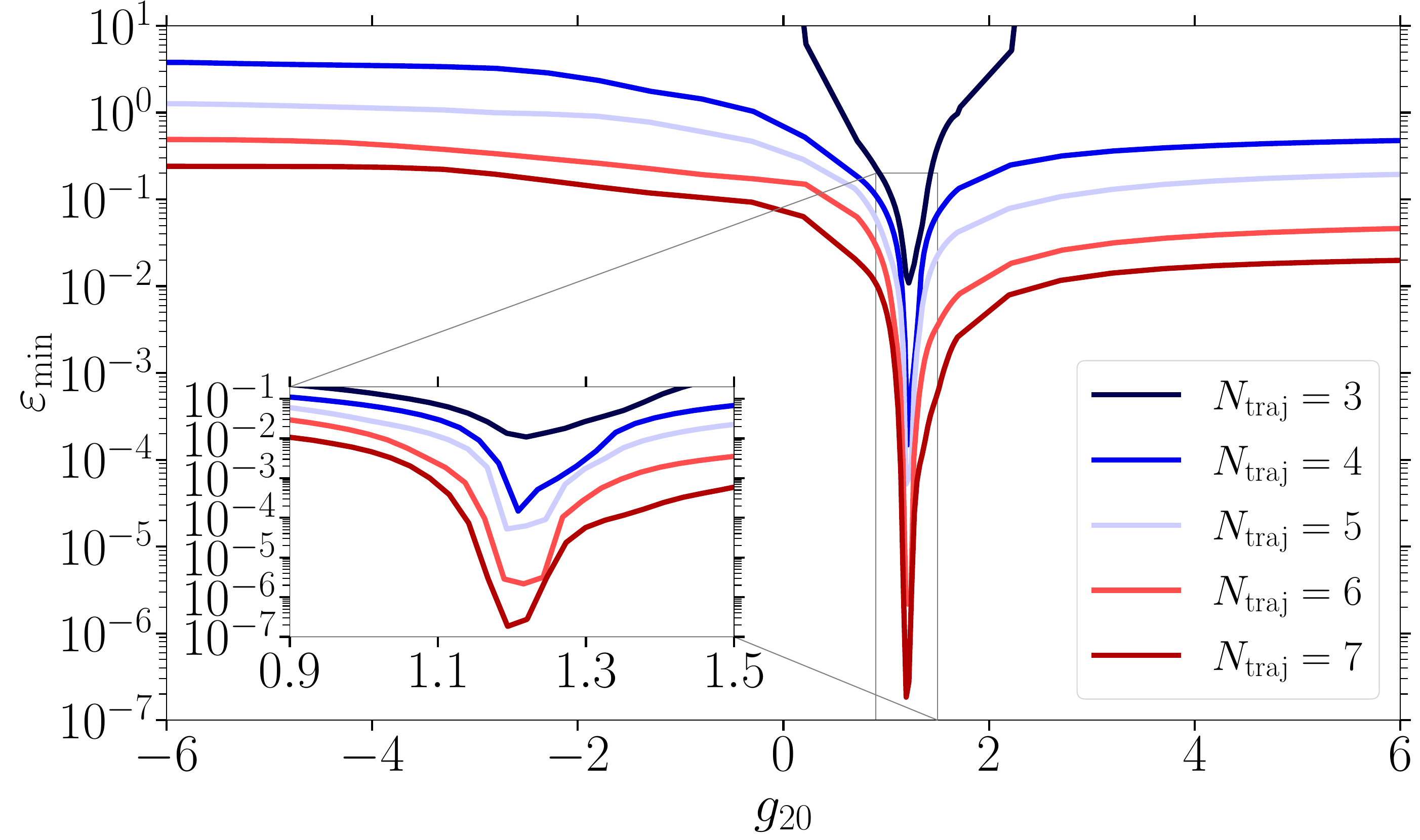}
    \caption{
     Veneziano linear trajectories ansatz. Slices through the $(g_{20}, g_{11})$-plane of fig.~\ref{fig:shrinking-ellipses-vene} in terms of the minimally allowed tolerance $\epsilon$ leading to feasibility of the imposed crossing constraint of eq.~(\ref{eq:crossingconstraints}) per number of trajectories $N_{\mathrm{traj}}$. We fix $g_{11}$ to the value attained for the global minimal $\epsilon$. The inset plot zooms into the region around the pit detected at the Veneziano amplitude's position.} 
     \label{fig:dip-vene}
\end{figure}

Figs.~\ref{fig:shrinking-ellipses-vene} and~\ref{fig:dip-vene} show essentially the same phenomenon. The first shows, for various \textit{fixed} tolerances $\epsilon$ the corresponding allowed space, at a fixed number of trajectories. It zooms exactly on the Veneziano amplitude. The second picture shows, for some fixed $g_{11}$ what is the minimum achievable $\epsilon$ for all values of $g_{20}$, with varying number of trajectories. We see that the tolerance globally converges, but that it converges even faster near the Veneziano value ($g_{20}\simeq 1.21$) and it displays a large dip at this value. This is one of the main results of this paper.

\subsection{Bending trajectories}
\label{sec:results-bent}
Having established how our bootstrap detects amplitudes, we now try to build amplitudes with non-linear trajectories.

In string theory, the extreme softness of the strings yields the famous exponential suppression at fixed angles, see a recent discussion in~\cite{Haring:2023zwu}. In QFT, it is expected that amplitudes behave like power laws, up to logarithmic corrections. Therefore the trajectories should flatten as $t<0$. To mimic this phenomenon, we built tunable trajectories where slopes are different at $s>0$ and $s<0$, given by
\begin{equation}
    \alpha_\lambda(s) =  \alpha_1 \lambda s +\alpha_2 (1-\lambda) h(s) s
\end{equation}
where $h(s)=e^s/(e^s+1)$ is a smooth step function that interpolates  between $-1$ and $0$. This is the model depicted in fig.~\ref{fig:bent-trajectories}. 
%

\begin{figure}
    \centering
    \includegraphics[width=\columnwidth]{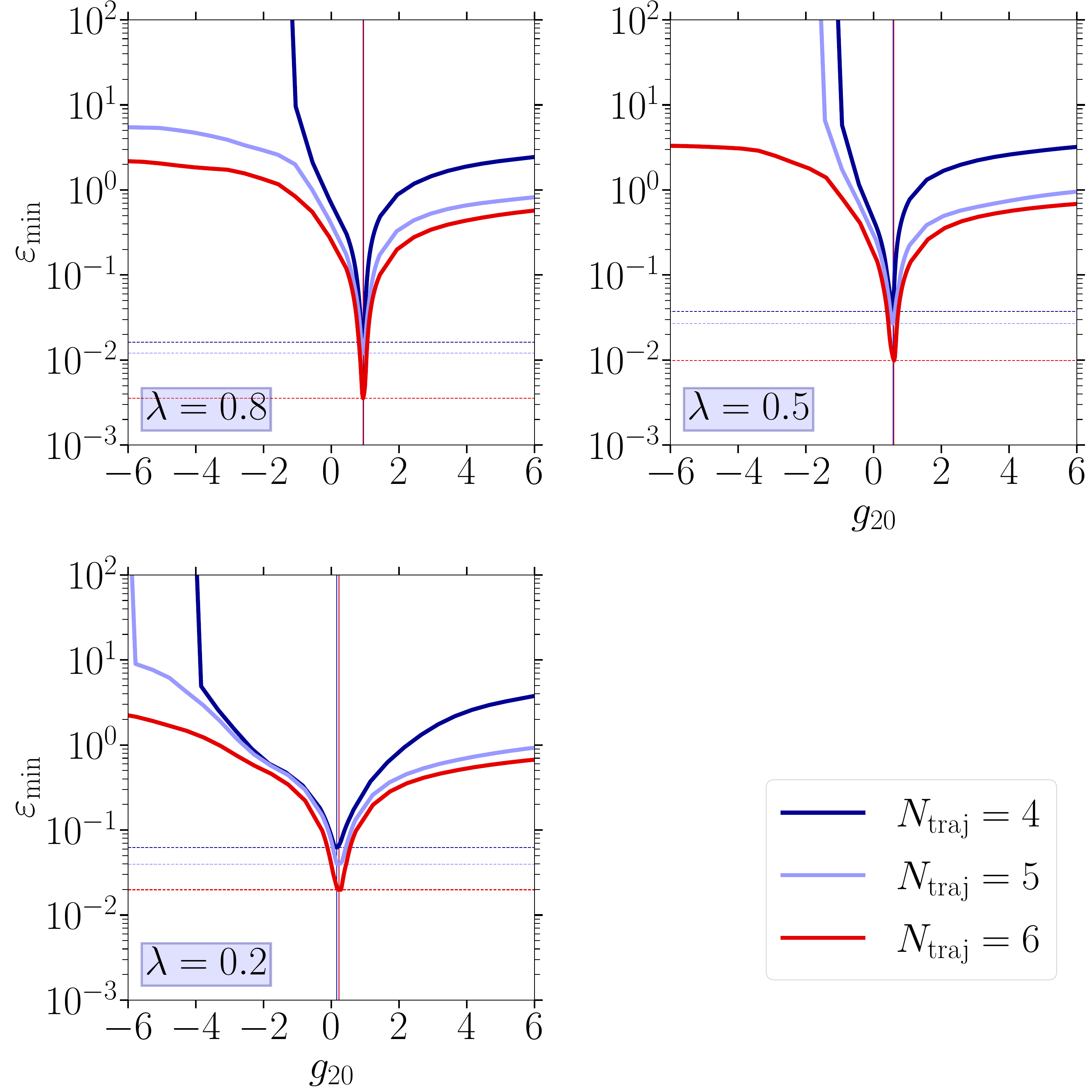}
    \caption{Model with bent trajectories. Slices through the $(g_{20}, g_{11})$-plane at fixed $g_{11}$ and increasing number of sub-leading trajectories showing the minimally allowed $\epsilon$ leading to feasibility of the the imposed crossing constraint of Eq.~\ref{eq:crossingconstraints}. We impose crossing for three values of $\lambda\in\{0.2, 0.5, 0.8\}$ dictating the degree of trajectory bending. For each $\lambda$ and $N$, $g_{11}$ is fixed to the numerically derived value attained when imposing the global minimal tolerance $\epsilon$ before infeasibility. The colored vertical lines -- adhering to the color choice for different $N$ -- illustrate the position of $g_{20}$ corresponding to the lowest $\epsilon_{\mathrm{min}}$ while the colored horizontal lines refer to the value of $\epsilon_{\mathrm{min}}$ per $N$.}
    \label{fig:bentTraj-dips}
\end{figure}

We first establish that such deformed amplitude are detected by our algorithm by a dip similar to Veneziano, which we show at different values of $\lambda$ in fig.~\ref{fig:bentTraj-dips}.\footnote{We do not show the analogue of fig.~\ref{fig:shrinking-ellipses-vene} because the plots bring no new information.}
This is the other most important plot of the paper, which shows that we can generate amplitudes with non-linear trajectories still located by a large dip, hinting at their existence.

\begin{figure}
    \centering
    \includegraphics[width=0.98\columnwidth]{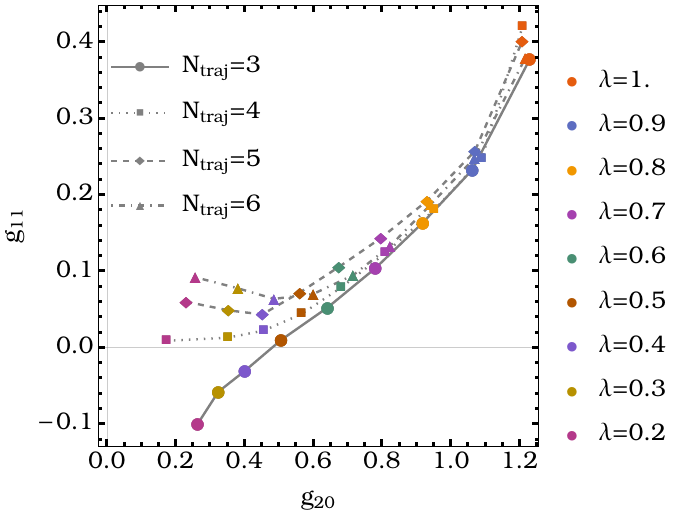}
    \caption{Evolution of the position of the dip at different values of $\lambda$ for $N_{\text{traj}}$ from 3 to 6.}
    \label{fig:lambda-flow}
\end{figure}

We then show how the amplitude moves in the $g_{11},g_{20}$-space as we vary $\lambda$ in fig.~\ref{fig:lambda-flow}. For strong bending ($\lambda\simeq 0.2$), our algorithm starts to lack precision. Nevertheless, the amplitudes follow an identifiable path, and it would be fascinating to explore the robustness of this evolution with respect to other similar bent deformations.

\begin{figure}
    \centering
    \includegraphics[width=0.95\columnwidth]{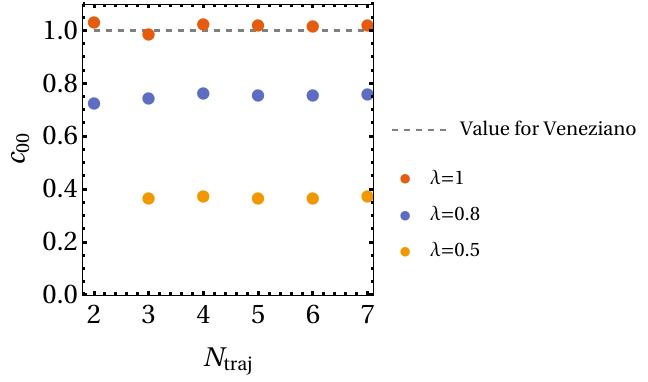}
    \caption{Convergence of the $c_{00}$ Gegenbauer coefficient for the bent amplitudes. The data point at lambda=0 and $N_{\rm traj}=2$ is missing because one needs at least $N_{\rm traj}=3$ to find amplitudes with such small $\lambda$.}
    \label{fig:bentc00}
\end{figure}

Finally in fig.~\ref{fig:bentc00}, we display how the coefficients $c_{n,n}$ of the leading trajectory converge with $N_{\mathrm{traj}}$. They clearly enough decrease with bending. At higher trajectories, our numerics loses in precision, see figs.~\ref{fig:bentci0}, but it seems that crossing forces the algorithm to shift the weight from the leading to more and more subleading trajectories. It would be interesting to understand this better, by including more trajectories and improving precision.

\section{Discussion}
\label{sec:discussion}

In this paper we have presented a numerical bootstrap for dual model amplitudes with linear and non-linear trajectories. It takes as an input the whole set of Regge trajectories, which allows us to render crossing constraining, enabling the detection of special amplitudes with large dips. Within the class of residues we studied (polynomially parametrized with a Veneziano-like global function), the Veneziano amplitude appears at such a dip. We could then proceed to deform it to non-linear trajectories and still maintain the dip.

One may ask if the data of the entire set of trajectories are not too much. After all, in the past and more recently, bootstraps aimed to \textit{determine} the Regge function, not use it as an input. In comparison, the conformal field theory (CFT) bootstrap~\cite{Poland:2022qrs,Hartman:2022zik} only needs data about a few operators, not the whole spectrum, to make crossing constraining.  We attribute this to the fact that in the CFT bootstrap, a region exists that lets the operator product expansion converge fast, unlike for the dual model bootstrap, where we have to combine several regions and use the whole spectrum. 

It would, of course, be desirable to be able to get constraints on the Regge function itself. This bootstrap allows for such a thing. Firstly, note that it is very much in the spirit of the Mandelstam iteration~\cite{Mandelstam:1958xc,Mandelstam:1959bc,Mandelstam:1963iyb,Atkinson:1968hza} recently used in~\cite{Tourkine:2023xtu}, where the multi-particle production is used as an input to parametrize the space of all non-perturbative amplitudes. Here, scanning the space of Regge trajectories, one scans the space of dual models, so that the problem is actually complete.

One immediate application would be to observe that amplitudes which violate linearity $\alpha(s) = s^\gamma$ with $\gamma>1$ at high energy are disfavored within this scheme, in accordance with~\cite{Caron-Huot:2016icg,Haring:2023zwu}. We leave this for future work.

\paragraph{Convergence}
To what extent can we say that our bootstrap allows us to locate \textit{existing} amplitudes? Firstly, we have observed a very clear signal of convergence of the tolerance parameter: on the dip plots of figs.~\ref{fig:dip-vene},\ref{fig:bent-trajectories}, the minimum achievable parameter $\epsilon$ decreases roughly with $c^{N_{\mathrm{traj}}}$. However, the same can be said of any point in the curve, since they are all globally shifted by the same constant.

The most prominent feature of these plots is the presence of a large dip, indicating a place where crossing plunges and can be satisfied by orders of magnitude better. It is not \textit{a priori }obvious how to interpret such a signal, even if very strong, but in combination with the fact that in the case of Veneziano we know that it correlates exactly with the presence of a special amplitude, we would like to conjecture that such dips marks the presence of special objects, amplitudes which satisfy ACU.
In the case of deformed, bent-trajectory amplitudes, the dip persists and this is why we want to conjecture that such amplitudes exist, too.

One interesting challenge, or puzzle, is however illustrated by the white ellipse in fig.\ref{fig:shrinking-ellipses-vene}. At a given number of trajectories, the crossing constraints we use only give a rather wide ellipse, if we apply the value at which the solution (the Veneziano amplitude) satisfies it. Yet, we observe that if we crank this tolerance factor beyond the Veneziano value, instead of ruling out the amplitude, we converge to it. The solution to this conundrum is maybe the key to improving our algorithm.

\paragraph{Numerics, Precision}
The technology used in this paper (linear programming, \textsc{mosek}), was geared towards speed. Because developing the bootstrap took us a long time, with a lot of trial and error, we needed speed above all. Yet, in this paper, we reach a decent precision of order 1\% for the location of the Veneziano $g_{20},g_{11}$ couplings. Convergence of other observables such as the $c_{n,J}$'s on subleading trajectories is less good. To improve, as we said, we need to add more trajectories. This cannot be done with our solvers, but in principle it can be done using SDPB\footnote{\url{https://github.com/davidsd/sdpb}}~\cite{Simmons-Duffin:2015qma,Landry:2019qug}, which would allow the possibility to do some precise measurements of the couplings $c_{n,J}$ of the amplitude we consider. 

\paragraph{Analytic structure of $\alpha(t)$.} Our ansatz function $h(t)$ has infinitely many poles in the complex plane. Thus, for some complex value of say $t$ near one of these poles $t_p$, at large $s$ the scattering amplitude would have a singularity $A(s,t)\sim s^{1/(t-t_p)}$. This does not appear desirable for a healthy meromorphic amplitude. It would be good to produce better models of Regge trajectories. This would probably go in parallel with understanding better the expected properties of the universal factor $B(s)$ and how to generalize it.

Note also that our approach to bending is alternative to that presented in \cite{Veneziano:2017cks} where trajectories intersect. It would be interesting to see if both approaches can be merged. 

\paragraph{Space of amplitudes with linear, integer spaced trajectories.} Our ansatz~\eqref{eq:beta-ansatz} does not describe the whole class of amplitudes with linear trajectories. The authors of~\cite{Haring:2023zwu} studied a shifted ansatz~\cite{Khuri:1969elr} and argued that it is a complete basis of amplitudes with linear trajectories and equidistant spectrum. In particular, one simple example, the Matsuda amplitude~\cite{Matsuda:1969zz}, is an amplitude for which the Regge residues have an extra $\lambda^s$ term multiplying the polynomials. Hence, our residue function is restricted, and can be improved. As a step-0 improvement, we have tried to enlarge the degree of the polynomials: this did not result in new dips, but it however turned the pointy spikes of the dips into slightly larger valleys, of bowl shape.

\paragraph{Relation to recent works.}
Meromorphic amplitudes have received increased attention following the work of \cite{Caron-Huot:2016icg} on the uniqueness of Veneziano asymptotics. In~\cite{Cheung:2023uwn}, a remarkable analytic construction of 4-point dual model amplitudes with non-linear trajectories was then found. Subsequently, it became apparent that a  portion of the resulting objects does not factorize correctly~\cite{Arkani-Hamed:2023jwn} and hence is ruled out by higher point factorization. What is remarkable here is that such a statement could even be made. Being able to use higher-point factorization in our approach would be essential too, but will require more work. At any rate, it would be very interesting to explore potential connections to our current work. 

Another alternative approach to dual models uses positivity and null constraints~\cite{Caron-Huot:2020cmc,Arkani-Hamed:2020blm,Bellazzini:2020cot,Tolley:2020gtv,Bern:2021ppb,deRham:2022hpx}. \textit{A priori}, this approach does not enable restricting to purely meromorphic amplitudes. Yet, for gauge theory amplitudes, the color factors may render it possible to select specific combinations which correspond to large-$N$ theories~\cite{Albert:2022oes}. In the recent follow-up~\cite{Albert:2023seb}, it was discovered that large-$N$ QCD might lie at a kink, and zooming in on the kink allowed the authors to produce spectacular spectra with linear trajectories and, in particular, obtain masses of mesons of a large-$N$ theory surprisingly matching those of $N=3$ QCD. A notable distinction between their approach and ours lies in the trajectory determination: in their work, the trajectory is identified, while in our method, it is an \textit{a priori} input. Investigating the potential complementarity of both approaches would be of significant interest.

Finally, a lot of works on the non-perturbative S-matrix bootstrap have seen the emergence of Regge trajectories of states~\cite{Bose:2020cod,EliasMiro:2022xaa,Acanfora:2023axz,Gumus:2023xbs,Guerrieri:2023qbg,Guerrieri:2021ivu,Guerrieri:2022sod} and it would be really interesting to make contact with these fully non-perturbative objects with our approach.

It would finally also be interesting to try to compute Coon-like dual models~\cite{Coon:1969yw,Figueroa:2022onw,Geiser:2022exp,Cheung:2022mkw,Cheung:2023adk}, maybe some that would closer resemble~\cite{Klebanov:2006jj,Maldacena:2022ckr} for applications towards string theory and holography.

\onecolumngrid
\noindent\rule{\linewidth}{0.1pt}

\noindent\textit{Acknowledgments.}
We thank 
Mehmet Gumus, 
Denis Karateev, 
Grisha Korchemsky, 
Kelian H\"aring, 
Miguel Paulos, 
Balt van Rees, 
Zhenia Skvortsov 
and 
Sasha Zhiboedov 
for very useful discussions and comments.
This project has received funding from Agence Nationale de la Recherche (ANR), project ANR-22-CE31-0017. 
The work of CE is supported by the ANR through grant ANR-19-CE31-0005-01 (PI: F.~Calore), and has been supported by the EOSC Future project which is co-funded by the European Union Horizon Programme call INFRAEOSC-03-2020, Grant Agreement 101017536. CE further acknowledges support from the COFUND action of Horizon Europe’s Marie Sk\l{}odowska-Curie Actions research programme, Grant Agreement 101081355 (SMASH).
\vspace{6pt}
\rule{\linewidth}{0.1pt}

\twocolumngrid

\appendix


\section{Definitions}
\label{app:definitions}
We follow the conventions of \cite{Collins_1977,Correia:2020xtr}.
The Mandelstam variables are defined by $s=-(p_1+p_2)^2=4E^2,\,t=-(p_1+p_4)^2 = -4E^2(1-\cos(\theta)),\,u=-(p_1+p_3)^2 = -4E^2(1+\cos(\theta))$, where $E$ is the center of mass energy of the process and $\theta$ the scattering angle. They satisfy $s+t+u=4m^2$.

In $d$ dimensions, the Gegenbauer $P$- and $Q$- functions are defined as
\begin{equation}
    P_j^{(d)}={}_2F_1(-J,J+d-3,\frac{d-2}2,\frac{1-z}2)\,.
\end{equation} 
and
\begin{multline}
Q^{(d)}_J(z)=\frac{\sqrt{\pi}\Gamma(J+1)\Gamma(\frac{d-2}{2})}{2^{J+1}\Gamma(J+\frac{d-1}{2})} \frac{1}{z^{J+d-3}}\times\\\times {}_2F_1\bigg(\frac{J+d-3}{2},\frac{J+d-2}{2},J+\frac{d-1}{2}, \frac{1}{z^2}\bigg)\ .
\end{multline}

\section{Mandelstam-Regge poles.}
\label{app:MRP}
\subsection{Bits of Regge theory}
\label{app:Regge}
Regge theory~\cite{Regge:1959mz} aims at understanding the structure of scattering amplitudes at high energy by studying it at complex angular momentum.
Here, we recall briefly some concepts and tools of this theory, and sketch how Mandelstam introduced the refinement that gives rise to our ansatz~\eqref{eq:truncatedMRexp}. For more details about Regge theory, we refer to the textbook~\cite[Chap.~2]{Collins_1977}.

It all starts from the partial wave expansion
\begin{equation}
    A(s,t)= 16\pi \sum_{J=0}^\infty(2J+1) A_J (s) P_J (z)
    \label{eq:PWE}
\end{equation}
where the $A_J$'s are the partial waves. By means of a Sommerfeld-Watson transform, backed by the Froissart-Gribov formula to define the partial waves at continued spins, one can deform the discrete sum over spins into a sum over the singularities in the complex $J$-plane of the partial waves, plus a left-over, background integral over $J$.

The singularities we consider for meromorphic amplitudes are called Regge poles, near which the partial waves behave as
\begin{equation}
    A_J(s) \underset{J\to\alpha(s)}\sim \frac{\beta(s)}{J-\alpha(s)}
    \label{eq:Regge-poles}
\end{equation}
so that Regge poles are actual poles of the partial waves.
The leftover background integral is usually obstructed at $\Re J=-1/2$ because for  $\Re J<-1/2$ the Legendre functions grow again. The assumption of a negligible background integral contribution is, thus, not valid anymore.

Mandelstam's refinement aims at fixing this: send the background integral arbitrarily to the left and capture all the Regge poles in the left half-plane. It uses the formula:
\begin{equation}
\frac{P_J(z)}{\sin{\pi J}}-\frac{1}{\pi}\frac{Q_J (z)}{\cos{\pi
    J}}=-\frac{1}{\pi}\frac{Q_{-J-1}(z)}{\cos{\pi J}}
\end{equation}
to relace $P$'s by $Q$'s in the partial wave expansion. Since the $Q$ functions decay at infinity, this supposedly renders the background integral smaller. Finally, the symmetry of $Q$ functions at half-integer points $Q_J(z) =Q_{-1-J}(z)$, $J\in1/2+\mathbb{Z}$ is used to remove unwanted singularities for $\Re J<0$. For the exact derivation of the Mandelstam-Regge representation, we refer to~\cite[Chap.~2.9]{Collins_1977} and the original paper of Mandelstam~\cite{MANDELSTAM1962254}.\footnote{The curious reader should be aware that a few trivial typo spoil the derivation, in particular no factor of $(-1)^J$ is needed multiplying the $Q$ function.} 

One important element is that Mandelstam assumes a symmetry of partial waves around half-integer points identical to that of the $Q$'s, whose validity has not been established. For the purposes of our paper, it turns out that empirically the Veneziano amplitude satisfies a very good match to the MR pole expansion, and so we believe that our deformations do, too.

\subsection{Match of Veneziano to MR}
\label{app:vene-MR-check}
\begin{figure*}
    \centering
    \includegraphics[width=0.42\linewidth]{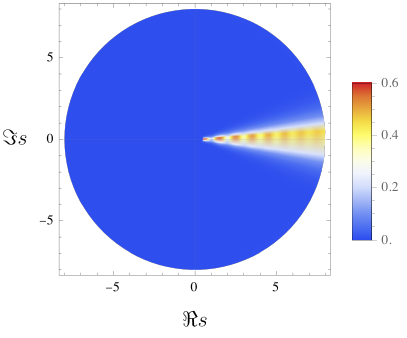}
    \qquad\quad\includegraphics[width=0.42\linewidth]{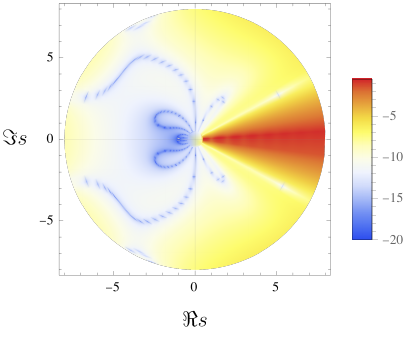}
    \caption{(\textit{Top}:) This displays $\bigg|1-\frac{A_{\rm Vene}^{N=8}(s,t)}{A(s,t)}\bigg|$ over the complex plane for $s$, with $\Im(s/t)=0$ with our $i \epsilon$ prescription and $|t|=2$, at $N=8$ trajectories. (\textit{Bottom}:) Logarithm of the same quantity. It shows that the MR expansion works extremely well away from the real axis, with the correct $i\epsilon$ prescription. We have no interpretations for the features of super convergence (blue Rorschach curves on the right plot).}
    \label{fig:MR-vs-Vene}
\end{figure*}

Since the validity of the ansatz is not proven in full generality, at many stages in this work we need to check things explcitly for the Veneziano amplitude. In fig.~\ref{fig:MR-vs-Vene} we display how well the MRPE approximates the Veneziano amplitude in the region of the $(s,t)\in \mathbb{C}^2$ plane with $\Im(s/t)=0$.

\subsection{Relation to dispersion relations}%
\label{app:regge-vs-DR}

In \eqref{eq:truncatedMRexp}, the quantities which parametrize the amplitude are $\alpha_i$ and $\beta_i$. To explain how we apply unitarity, we need to understand how they are related to the usual functions $m_n^2$ and $c_{n,J}$. This is most easily done when considering instead of the MRPE, the vanilla Regge pole expansion, for which 
\begin{equation}
\label{eq:Regge-expansion}
 A^{N_{\text{traj}}}(s,t)\sim 16 \pi^2 \sum_{i}(2\alpha_i(s)+1)\beta_i (s)\frac{P_{\alpha_i (s)}(-z)}{\sin \pi \alpha_i (s)},
\end{equation}
in terms of the same functions $\alpha$ and $\beta$. The difference between \eqref{eq:truncatedMRexp}
 and \eqref{eq:Regge-expansion} is that the latter has much better convergence properties and approximates the amplitude most efficiently.\footnote{Here and everywhere else, these sorts of statements are backed by empirical observation on the explicit Veneziano amplitude which we performed.}

It is now easy to see that when $\alpha(s)=n\in\mathbb{N}$ the amplitude develops single poles, i.e.~the masses of the resonances are defined by $\alpha(m_n^2)=n$. For the Regge residue $\beta$ we have that near a pole $\alpha(s)=n$, the amplitude behaves like 
\begin{equation}
    A(s,t)\simeq \sum_i 16\pi^2 (2(n-i)+1)\beta_i(n) P_n(1+2t/(4m_n^2-m_0^2), 
\end{equation}
which allows one to read off that 
\begin{equation}
    16 \pi ^2 (2(n-i)+1)\beta_i(m_n^2) = c_{n,n-i}.
\end{equation}

\subsection{Regge residues of the Veneziano amplitude}
\label{app:vene-residues}
We extracted the Regge residues of the Veneziano amplitude by using the Froissart-Gribov formula close to a Regge pole.\footnote{We thank S. Zhiboedov for a useful discussion on this point.} This provides an easy and efficient systematic way of computing Regge residues, when the amplitude is known.

Below we display the Regge residues up to the third trajectory, as it is the first non-trivial example.
We define first reduced coefficients
\begin{equation}
    c_{n,J} = h^{(d)}_{nJ} \tilde c_{n,J},
\end{equation}
where 

\begin{equation}
    h^{(d)}_{nJ}=\frac{n^n (d+2 J-3)\sqrt{\pi }  2^{-d-2 n+3} \Gamma (d+J-3)}{\Gamma \left(\frac{d-2}{2}\right) \Gamma (J+1) \Gamma
   \left(\frac{d-1}{2}+n\right)}
\end{equation}
We then have that:
\begin{equation}
\begin{aligned}
    \tilde c_{n,n} &= 1\\
    \tilde c_{n,n-1} &= \frac{-3+d+2n}{n}\\
    \tilde c_{n,n-2} &= \frac{(-3+d+2 n) \left(n^2+n(9-d)+2 d-10\right)}{6 n^2}\\
    \dots
    \end{aligned}
\end{equation}
Looking at the polynomial in the numerator of $\tilde c_{n,n-2}$, we observe that the shifted polynomial where $s\to s+2$, given by $12 + 12 d + (37 + 12 d - d^2) s + (27 - d) s^2 + 2 s^3$, has only positive coefficients for $d\leq14$. Up to the 37-th trajectory which we computed, we observed this property, which lead us to conjecture the super-unitarity of the Veneziano amplitude. This calculation extends that of~\cite{Maity:2021obe} to orders as high as one can compute, but does not provide a proof of the unitarity of the Veneziano amplitude directly from the Beta function yet.

\begin{equation}
    \begin{aligned}
        c_{0,0} &= a_{00}, \\
        c_{1,1} &= a_{00}/2, \\
        c_{1,0} &= \frac{a_{10}+a_{11}}{2 (d-1)}, \\
        \dots\\
        c_{n,n-j} &=h^{(d)}_{n,n-j}\frac{1}{n^{d_j}}\sum_{k=0}^{q_j} a_{jk}n^k  
    \end{aligned}
    \label{eq:a-to-c}
\end{equation}
\section{Details on crossing}
\label{app:crossing}

\subsection{Choice of the crossing regions}
\begin{figure*}
    \centering
    \includegraphics{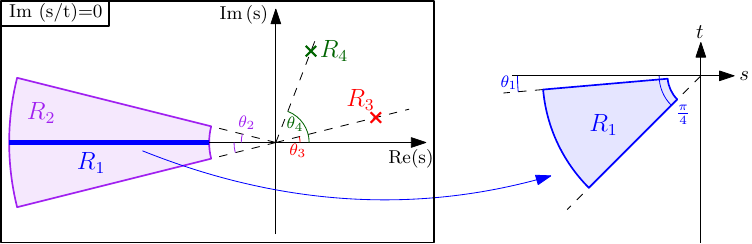}
    \caption{Depiction of domains of $(s,t)\in\mathbb{C}^2$ where we enforce crossing. Left: complex plane for $s$, where $t$ is constrained to lie on the same ray as $s$. Right: region $s,t<0$. We have: $\theta_1=2/10 (\simeq \pi/15),\theta_2\simeq\pi/12,\theta_3=1/4\simeq\pi/12,\,\theta_4=6/5\simeq3\pi/8$. We only use $R_4$ in appendix~\ref{app:sign-flip} to impose a non-trivial new sign constraint that gives a seemingly local minimum different from the Veneziano amplitude.}
    \label{fig:schematic-constraints}
\end{figure*}

Here we explain how to choose the set of points where crossing is imposed to obtain the results of the main text. First, let us stress that for a given truncation $N_{\text{traj}}$ there will be a minimal value of $\epsilon$ attainable by physical amplitudes, and enforcing a value of $\epsilon$ too small will cause their exclusion. Naturally, this extremal value of $\epsilon$ becomes smaller the more trajectories we add, and understanding how it evolves with $N_{\text{traj}}$ plays an important role when identifying physical amplitudes in our approach.

\begin{figure}
    \centering
    \includegraphics[width=0.8\linewidth]{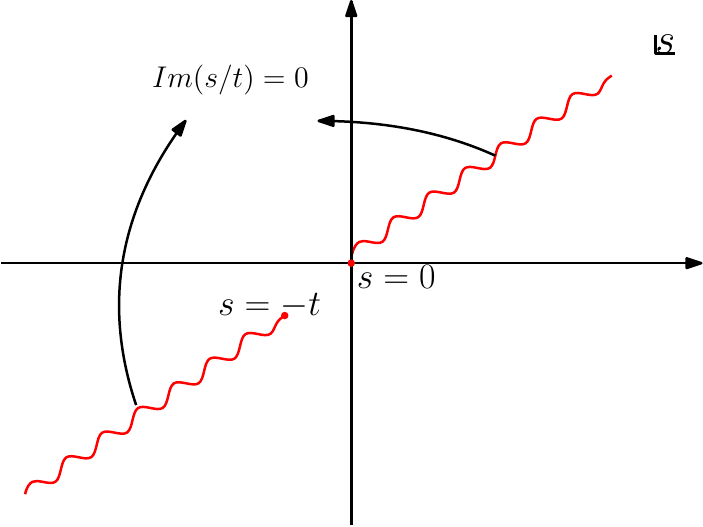}
    \caption{Singularity structure of $Q_{-s-1}(-1-2t/s)$ for fixed $t$.}
    \label{fig:qj-singularities}
\end{figure}

To understand in which regions of $(s,t)$ crossing can be imposed,  we must look at the analytic structure of the $Q_J(z)$ functions. For integer $J$, they have a branch cut extending from $z=-1$ to $z=1$, and allowing for non-integer $J$ opens a new discontinuity from $z=-1$ to $-\infty$. In both cases the $Q_J$'s have logarithmic singularities at the branch points $z= \pm 1$. Moreover, as a function of complex $J$ they have simple poles every time $J$ hits a negative integer. This is shown in the $s$-channel (for fixed $t$) in  in the case of linear trajectories in fig.~\ref{fig:qj-singularities}.

In the MR expansion, the $Q_J$'s dependence on $s$ and $t$ is always of the form $Q_{-\alpha_i (s)-1}(-1-2t/s)$ (and the converse in the $t$-channel expansion), and thus we see that there are branch cuts opening at $s=-t$ and $s=0$ and extending to infinity along the direction $\Im(s/t)=0$. Moreover, there are simple poles everytime $\alpha_i (s)$ or $\alpha_i (t)$ are non-negative integers.

This should raise some alarms, as the objects we are trying to build are meromorphic functions, and thus they cannot have branch cuts or logarithmic singularities. In particular, we do not expect any discontinuity in the amplitude when traversing these branch cuts, and therefore the MR expansion must stop being a good approximation for the full amplitude as one crosses them. One can indeed check this for the Veneziano amplitude: In terms of $z$ the MR expansion can approximate the amplitude in the lower half plane $\Im z\leq0$, and the match between the expansion and the full amplitude becomes increasingly better as one approaches $\Im z=0$ from below, where the $Q_J$'s branch cut lies. However, as one crosses the real axis and passes into the $\Im z\geq0$ region the discontinuity of the $Q_J$ functions causes the MR expansion to completely break down. This teaches us that there is a ``good side'' of the branch cut where we need to stay.

However, at the time of imposing crossing we need both $z_s=1+2t/s$ and $z_t=1+2s/t$ to satisfy $\Im (z_{s,t})\leq0$. This implies $\Im (t/s)\leq 0$ and $\Im (s/t)\leq0$, which clearly can be satisfied only for points obeying $\Im (s/t)=0$. Moreover, the convergence of the MR expansion worsens when $1+2t/s$ approaches $\pm1$ (or $1+2s/t$ approaches $\pm1$ for the $t$-channel) and one gets near the logarithmic singularities of the $Q_J$ functions. 

Finally, the $(\cos{\pi \alpha_i})^{-1}$ terms in the MR expansion introduce an infinite number of spurious poles at half-integer values of the $\alpha_i$'s. This issue has been a matter of concern and some degree of confusion in the old days leading to efforts to find residue functions that vanish at every half integer so as to cancel these poles. While this could be an extra constraint to impose on the $\beta_i$'s, we chose a somewhat simpler approach inspired by the Veneziano amplitude. In this case, the residue functions are such that they cancel only the negative half-integer poles, but the poles caused by the cosines at positive values of the Mandelstam variables are still present, spoiling the validity of the expansion in the positive real axis. However, as one departs even a little from the real line the expansion quickly becomes spectacularly accurate again. In our ansatz we mimic this behavior, imposing the cancelation of the negative integer poles by adding a suitable term to the Regge residues but allowing for the positive ones, staying within the class of Veneziano-like objects.

Taking into account these characteristics of the analytic structure of the components of the MR expansion, we chose to impose crossing on a grid of complex values of $s$ and $t$ such that $\Im (s/t)=0$, far enough from the singularities of the $Q_J$ functions and the positive real axis. In practice, we use a regular grid of real, negative values of $s$ and $t$ and a small number of points with non-zero imaginary part, checking the convergence of our results under the addition of new points. Given that the value of the $Q_J$ functions is ill-defined exactly on the branch cut, we use an $i \epsilon$ prescription to ensure that the MR expansion on both channels remains on the good side of the cut. A schematic summary of the regions we consider in our bootstrap approach is given in fig.~\ref{fig:schematic-constraints} while the explicit definitions are stated in the main text.

The details on the grid of points we used are as follows: For R1 we used a set of 767 points from radius $\rho=1$ to $\rho=5$ and angle $\theta=2/10$ to $\theta=\pi/4$ (using polar coordinates in the negative, real ($s,t$) quadrant given by $s=-\rho \cos \theta$, $t=-\rho \sin \theta$). For R2 we used a set of 270 points lying on rays $\Im (s/t)=0$ (namely $s=|s| e^{i \theta}$, $t=|t| e^{i \theta}$, with $|s| \in [1,8]$ and fixing $|t|=2$, and $\theta \in [2.9,3.4]$.

\begin{figure}
  \label{fig:ellipses-sliced-vene}
    \centering
    \includegraphics[width=\columnwidth]{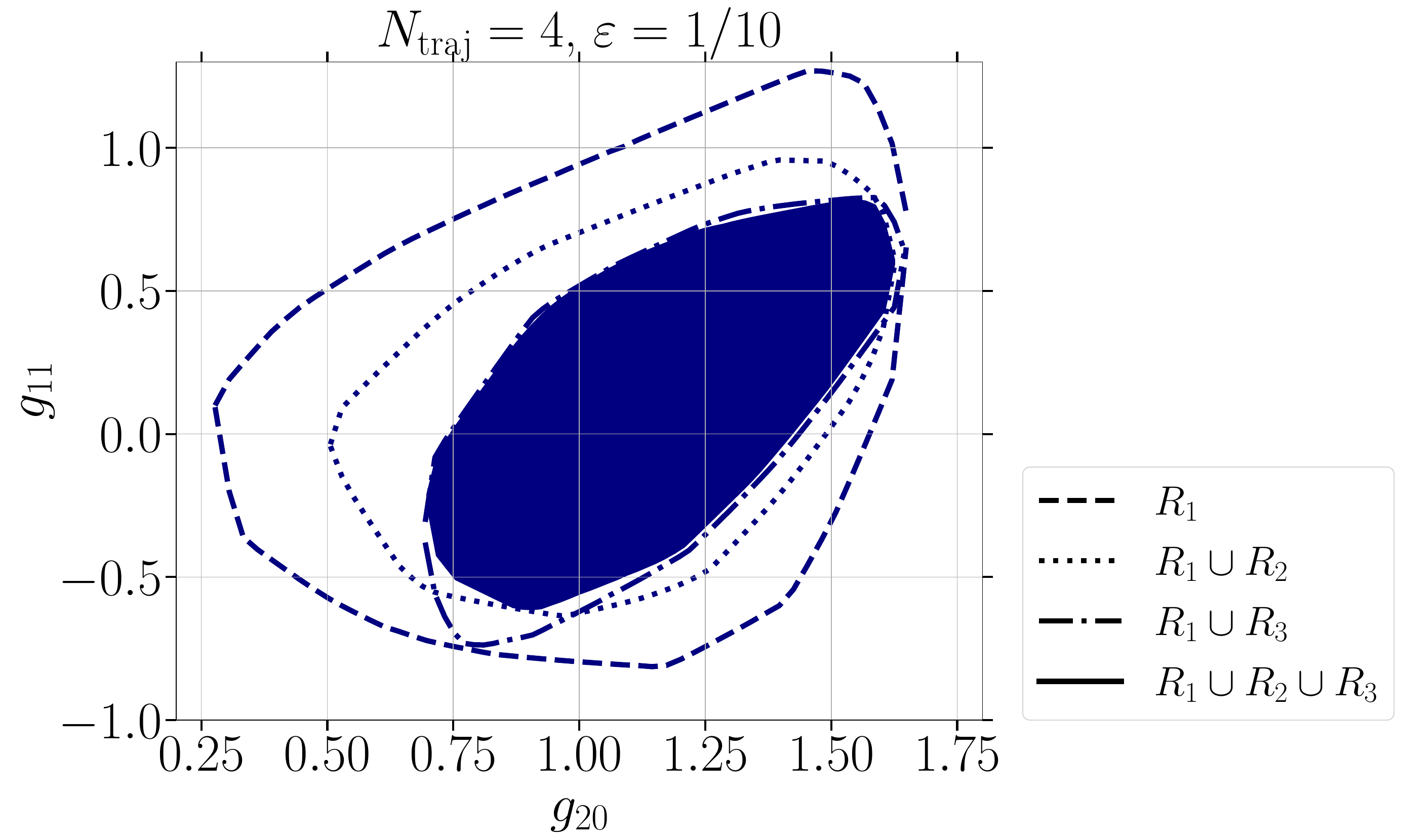}
    \caption{Allowed space in the $\left(g_{20}, g_{11}\right)$-plane evaluated at the point $(s, t) = (-5/3, -5/3)$ for a tolerance of $\epsilon = 1/10$ and a Regge ansatz featuring $N_{\mathrm{traj}}=5$ trajectories. We display with distinct line styles the allowed parameter space for different sets of points in the $(s,t)$-plane $R_i$ as illustrated in fig.~\ref{fig:schematic-constraints} imposed as constraints for our bootstrap while adhering to the sign convention obtained from evaluating the Veneziano amplitude at these points. The set of points $R_1\cup R_2\cup R_3$ -- the most constraining one -- represents the benchmark collection of constraints used for the results shown in the main text.}
\end{figure}

\subsection{Example of solutions via sign change in the positive region.}
\label{app:sign-flip}

\begin{figure*}
    \centering
    \includegraphics[width=0.49\linewidth]{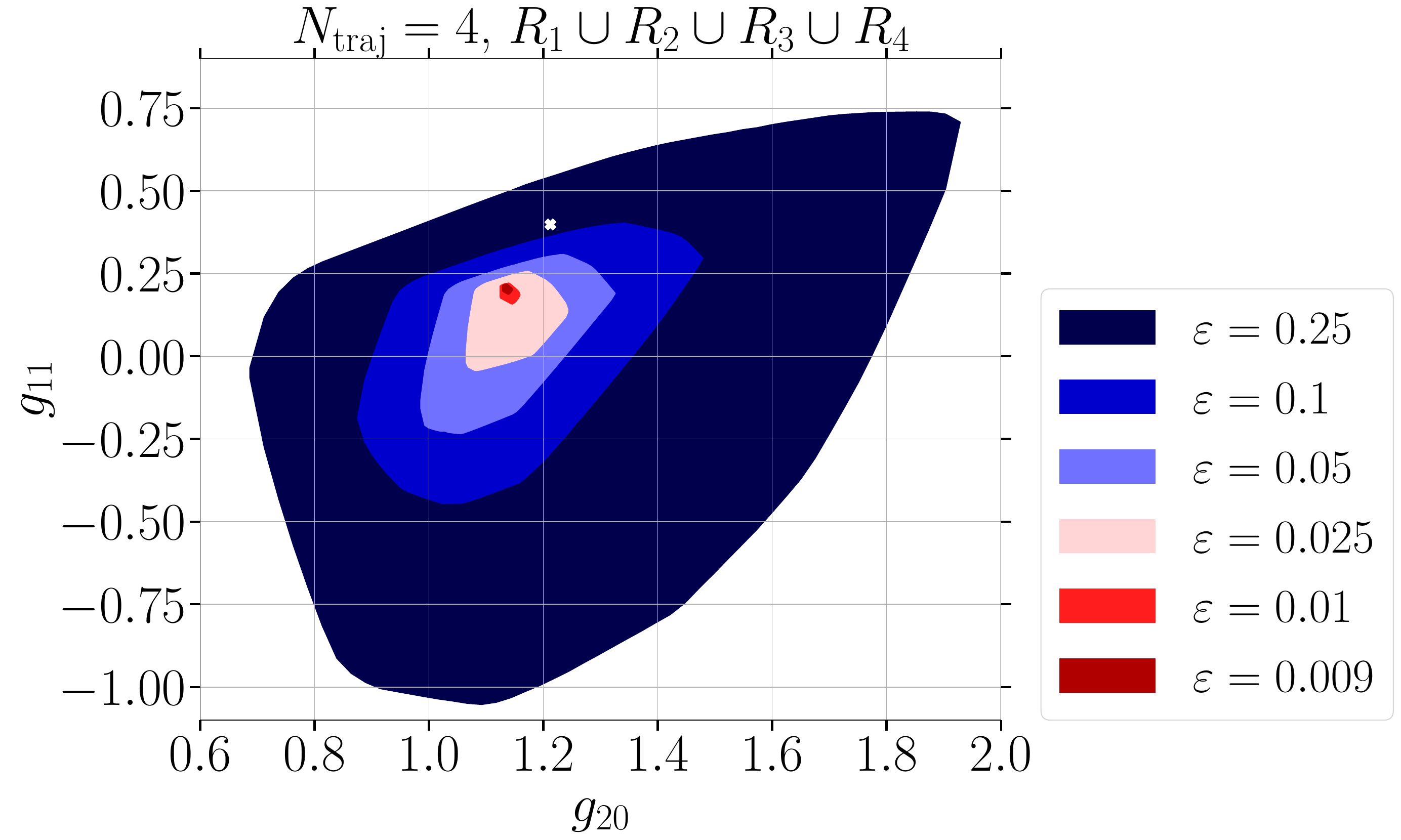}\hfill
     \includegraphics[width=0.49\linewidth]{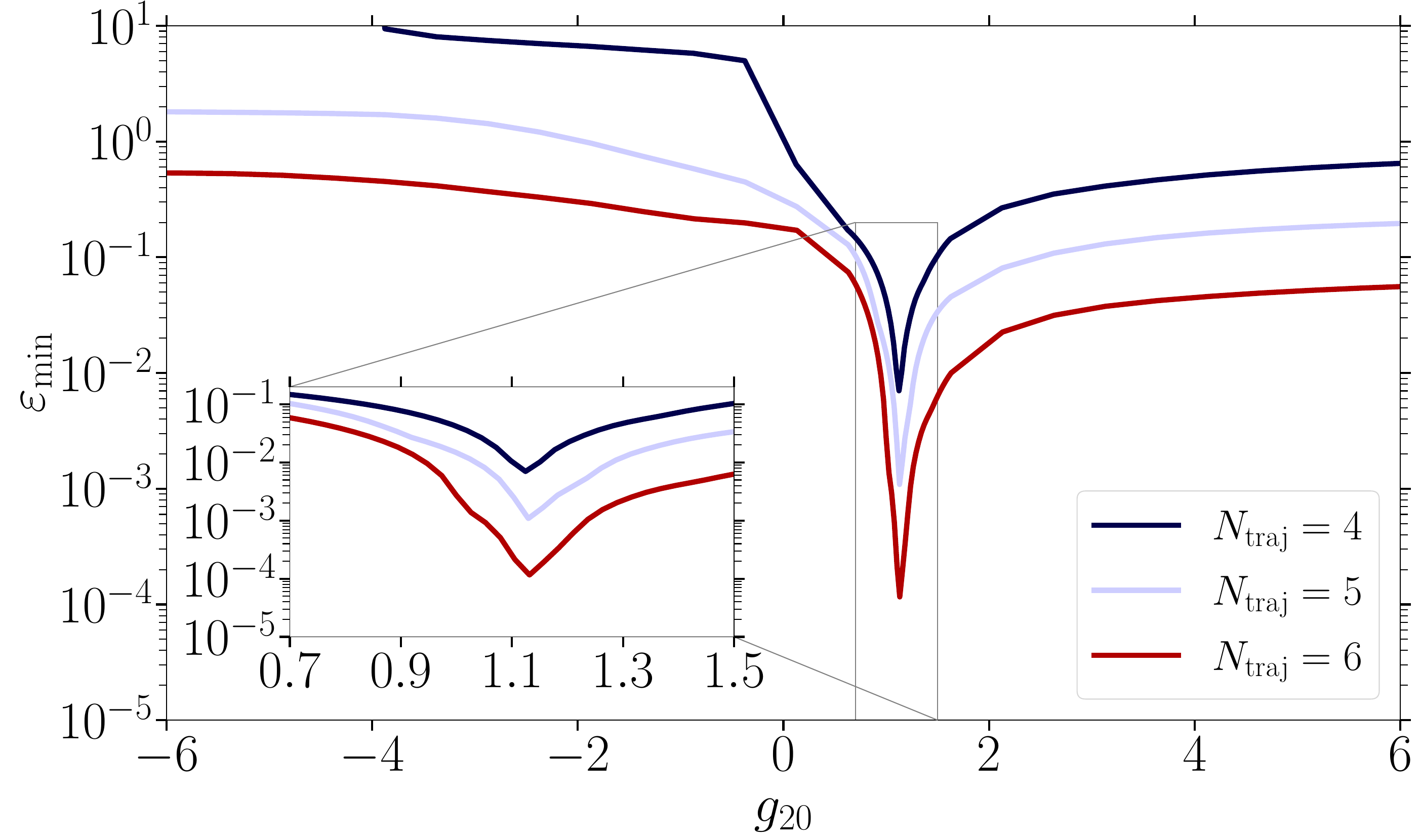}
    \caption{(\emph{Left}:) Same as fig.~\protect\ref{fig:shrinking-ellipses-vene} fixing the set of constraints to our benchmark setup and adding the point $R_4$ (see fig.~\ref{fig:schematic-constraints}) with the opposite sign that the Veneziano amplitude attains. We trace the allowed $\left(g_{20}, g_{11}\right)$ parameter space for decreasing values of tolerance $\epsilon$. The location of the Veneziano amplitude is marked with a white cross. We clearly find that the optimal solution regarding crossing symmetry shifts to a different location. (\emph{Right}:) Slices through the $(g_{20}, g_{11})$-plane of the left panel in terms of the minimally allowed tolerance $\epsilon$ leading to feasibility of the imposed crossing constraint of eq.~(\ref{eq:crossingconstraints}) per number of trajectories $N_{\mathrm{traj}}$. We fix $g_{11}$ to the value attained for the global minimal $\epsilon$.}
    \label{fig:eft_ellipses}
\end{figure*}

Finding the Veneziano amplitude as the optimal solution to our bootstrap approach resulted from strictly adhering to all of its known properties. This encompasses, among others, the sign of the Veneziano amplitude's real part for the set of constraints $R_1\cup R_2 \cup R_3$. Changing the sign of a single point of the constraints can be considered a minimal change of the ansatz; yet, it allows us to obtain a different minimum than the Veneziano amplitude as shown in fig.~\ref{fig:eft_ellipses}. To derive this new optimal solution, we add another point in the quadrant of positive real and imaginary part of $s$ and $t$ called $R_4$ at $(s, t) = (9/2e^{6i/5}, 2e^{6i/5})$ (see fig.~\ref{fig:schematic-constraints}). For this point we impose the opposite sign of what is obtained for Veneziano. The discovered solution behaves in some sense comparable to the Veneziano amplitude as it is characterized by a pit in the $(g_{20}, g_{11})$-plane around its position per $N_{\mathrm{traj}}$ (see the right panel of fig.~\ref{fig:eft_ellipses}). A quantitative difference between this solution and the Veneziano amplitude is the evolution of the minimal tolerance $\epsilon$ with $N_{\mathrm{traj}}$, which decreases slower in this example. Nonetheless, the characteristic pit lends credence to the fact that our bootstrap method has successfully uncovered a valid object. Yet, extracting its properties requires more adequate numerical tools than we currently have at our disposal and is, thus, left for future work.

\begin{figure*}
    \centering
    \includegraphics[width=0.49\linewidth]{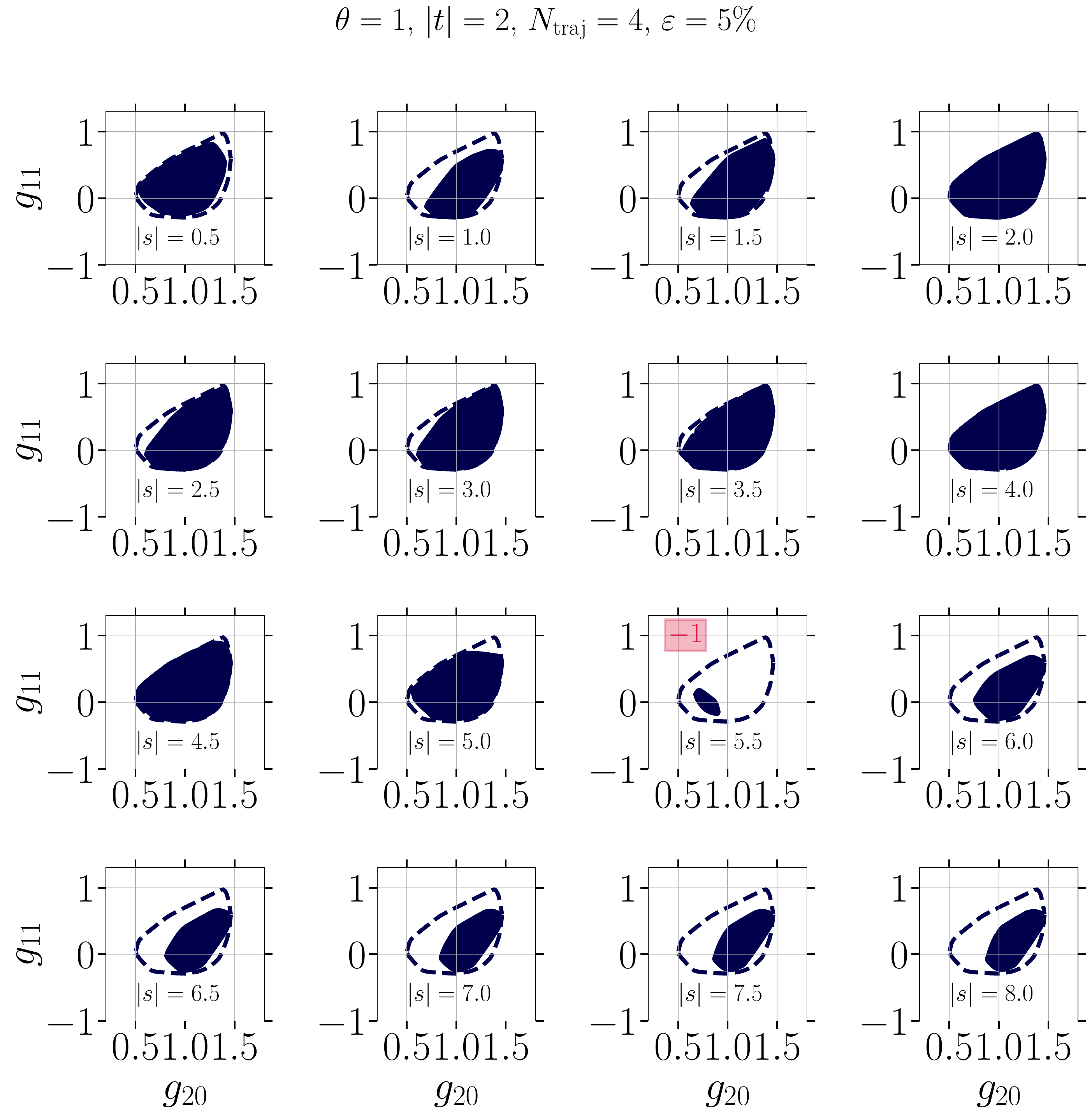}\hfill
     \includegraphics[width=0.49\linewidth]{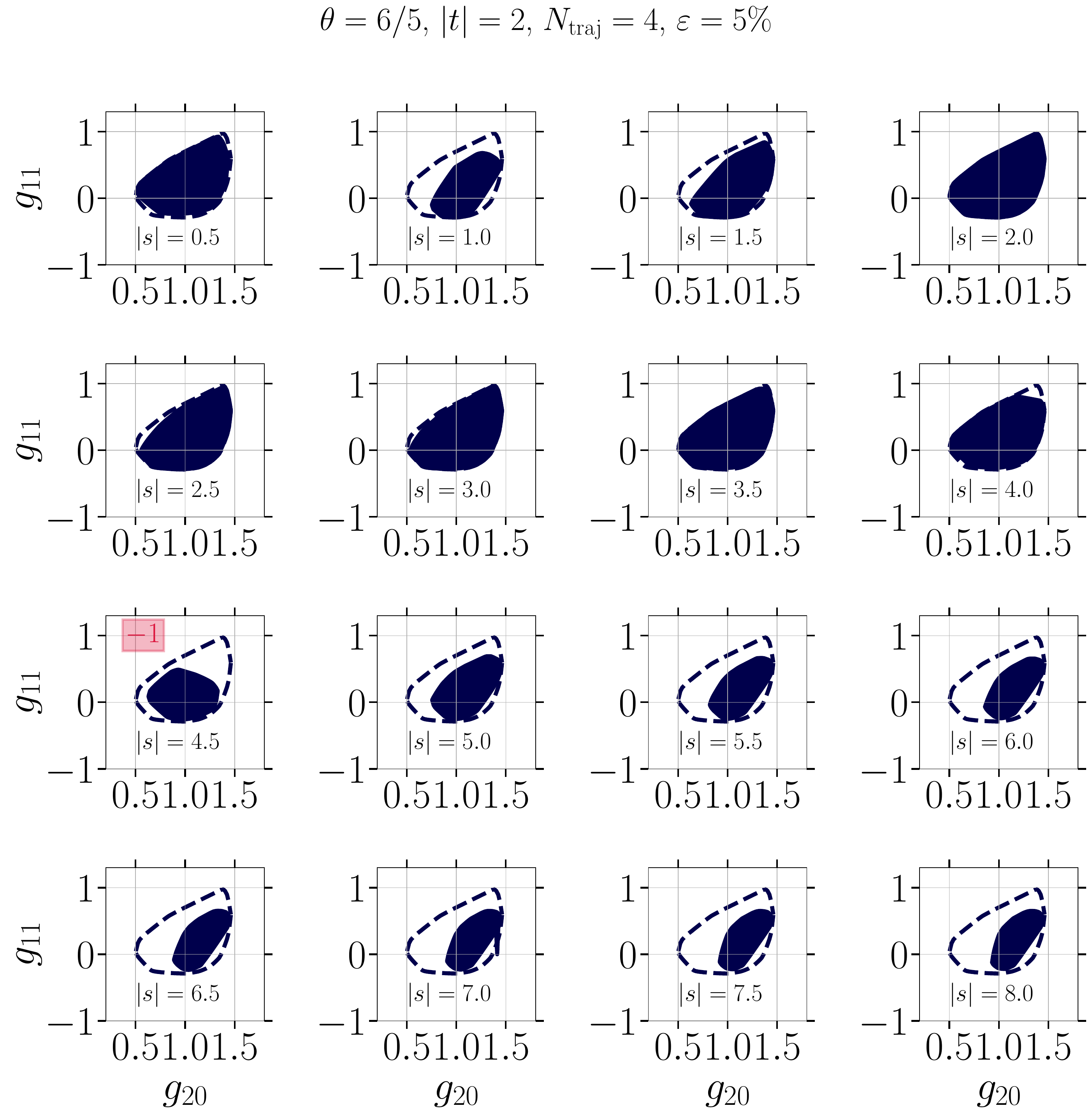}
    \caption{Scan over a set of points with $\Im(s/t) = 0$ and $\Re s, \Re t, \Im s, \Im t > 0$ similar to the results of fig.~\ref{fig:ellipses-sliced-vene}. We fix $N_{\mathrm{traj}} = 5$ and the tolerance parameter to $\epsilon 0 5\%$ to track the allowed parameter space in the $\left(g_{20}, g_{11}\right)$-plane when imposing crossing constraints from $R_1$ (dashed) and $R_1\cup R_3$ (solid). The set $R_3$ consists of a single point defined by $(s,t) = (|s|e^{i\theta}, 2e^{i\theta})$. The modulus of $s$ is reported in each panel, respectively. Panels marked with a red ``$-1$'' indicate a departure from the sign convention dictated by the Veneziano amplitude. (\emph{Left}:) $\theta = 1$.  (\emph{Right}:) $\theta = 6/5$.}
    \label{fig:ellipse-carving}
\end{figure*}

As a more general note, applying a different sign convention for point $R_4$ allowed us to satisfy crossing to a smaller degree than how the Veneziano amplitude performs. This is exactly the way we selected this particular point. Examples of a larger scan of points in this region of the $(s,t)$-plane is provided in fig.~\ref{fig:ellipse-carving} where we display instances of the left panel of fig.~\ref{fig:ellipses-sliced-vene} for different choices of $R_3$ and fixed $\epsilon = 1/20$. The selected examples lie on complex rays $\Im(s/t) = 0$ for varying $|s|\in\left[1/2, 8\right]$, fixed modulus $|t| = 2$ and two angles $\theta \in \{1, 6/5\}$. 

In these plots, we illustrate the impact of adding such points $R_3$ to the set of constraints from $R_1$ alone. The allowed parameter space in the $(g_{20}, g_{11})$-plane for a tolerance of $\epsilon = 5\%$ and $R_1$ constraints is displayed as dashed line whereas the filled ellipse denotes the allowed parameter space when imposing $R_1\cup R_3$. We observe as a general trend that points at large separation $|t| - |s|$ are more constraining while the shape of the remaining allowed parameter space does not exhibit much variation. This statement applies to the majority of other angles $\theta$ (not shown here) that we probed in the course of this work.

Remarkable exceptions to this trend are the two panels marked with a red ``$-1$''. Here, we depart from the Veneziano amplitude's sign convention that is assumed in all other panels. Doing so, drastically alters the shape of the $R_1\cup R_3$ parameter space that is incompatible with the Veneziano amplitude. The point belonging to this specific plot in the right panel of fig.~\ref{fig:ellipse-carving} is exactly $R_4$ used to derive the solution presented above. The other point in the left panel is at the same incompatible with Veneziano and the other minimum. Hence, it represents yet another object with linear trajectories and integer spacing belonging to the class of Veneziano-like amplitudes. We therefore conclude that our bootstrap approach is capable of numerically uncovering new objects with only a minimal change of the ansatz.

\section{Numerics}

For the results presented in this article, we chose to remain within the realm of linear programming, for which we employ the commercial solver \textsc{mosek}, under free academic license, utilizing its implementation in \textsc{mathematica} and a \texttt{python} interface provided by the \textsc{cvxpy} package \cite{diamond2016cvxpy, agrawal2018rewriting}. As an alternative, we explored and used the functionalities of ``SoPlex'' \cite{soplex-cite}; in particular, its capability to derive exact solutions based on rational numbers. 

During the development, we tried many numerical strategies, including quadratic/conic optimization. The quadratic/conic problems corresponded to a class of problems where we minimize crossing, instead of mini-/maximizing observables. For instance, we would define an objective function as $\mathcal{O}=\sum_{i\in\rm{points}} (A^N(s_i,t_i)-A^N(t_i,s_i))^2$ and minimize it. 

Compared to the current implementation based on uniform ratios as in eq.~\eqref{eq:crossingconstraints}, minimizing the crossing equation itself turned out to come with major drawbacks: \textit{(i)} Even though a conic formulation of a quadratic problem improves the numerical performance of the solver, round-off errors can prevent the algorithm from initializing properly (spurious negative eigenvalues); \textit{(ii)} Imposing crossing in terms of the objective function above with measure 1 introduces many useless constraints whenever and wherever the MR ansatz decays exponentially. In particular, the region $R_1$ for large $|s|, |t|$ does not add much information to the problem; \textit{(iii)} Along the same lines, adding points in $R_3$ is very delicate since the MR ansatz grows exponentially with $s$ and $t$. Picking suitable points leads to a very tailored problem without sufficient flexibility.    

Eventually, our strategy does essentially the same thing, while remaining within linear programming, hence being much faster. 

The main limitation of \textsc{mosek} is that it only works with double point precision, which for us means that $N_{\mathrm{traj}} = 7$ is the last reliable number of trajectories. SoPlex allows us to extend this limitation to an even larger number of trajectories of $N_{\mathrm{traj}} \sim 10$. In contrast to \textsc{mosek}, SoPlex exhibits a substantial increase in computation time for such numbers of trajectories ($\sim 8$h for ten trajectories) rendering quick exploratory work nearly infeasible. While it technically enables to work with arbitrary precision via rationalizing the constraints, we experienced SoPlex to struggle with either a large number of parameters or small tolerances $\epsilon < 10^{-7}$ that both inevitably occur for large $N_{\mathrm{traj}}$. 
The problem seems to be connected to the fact that some internal computations are still based on floating point numbers, which are rationalized in each iteration step.

Finally, we routinely used SDPB during the course of development of the project, but the large number of constraints which we process induced a slow down in speed not adapted to the exploratory work. 
For the future, if we want to bootstrap more precisely specific amplitudes, we will need to include more trajectories and hence work with arbitrary precision. For this, and in view of the issues with SoPlex, SDPB will be a very serious option.

\section{Higher trajectory coefficients with bending.}
\begin{figure*}
    \centering
    \includegraphics[width=0.46\linewidth]{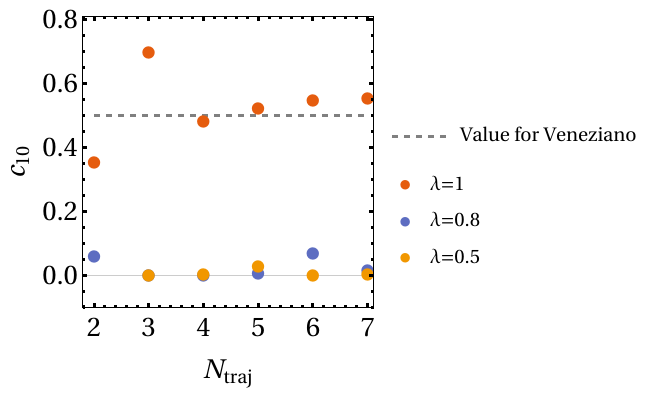}\qquad
    \includegraphics[width=0.49\linewidth]{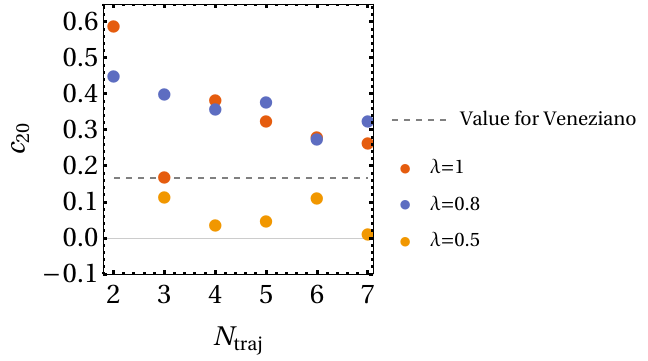}
    \caption{Left to right, top to bottom: coefficients $c_{10},c_{20},c_{30},c_{40}$ at various bending angles.}
    \label{fig:bentci0}
\end{figure*}

In this final appendix, we display in fig.~\ref{fig:bentci0} some not well converged amplitudes, just to give a few more observables about our amplitudes and demonstrate that they are all fully explicit. In principle using arbitrary precision solver we can push to higher number of trajectories and converge better.

\onecolumngrid

\newpage

\bibliographystyle{jhep}
\bibliography{biblio.bib}

\end{document}